\documentstyle[buckow]{article}

\def\half{{\textstyle {1 \over 2}}}
\def\quart{{\textstyle {1 \over 4}}}
\def\noverm#1#2{{\textstyle {#1 \over #2}}}
\def\Dpartial{{\cal D}}
\def\tr{\,{\rm tr}\,}
\def\unitmatrixDT%
 {\maththroughstyles{.45\ht0}\z@\displaystyle {\mathchar"006C}\displaystyle 1}
\begin {document}
\begin{flushright}
\small
UG/00-17\\
VUB/TENA/00/09\\
{\bf hep-th/0011264}\\
November 29, 2000
\normalsize
\end{flushright}
\def\titleline{
On the supersymmetric non-abelian Born-Infeld action
}
\def\authors{
E.~A.~Bergshoeff\1ad, M.~de Roo\1ad and A.~Sevrin\2ad
}
\def\addresses{
\1ad Institute for Theoretical Physics\\
Nijenborgh 4, 9747 AG Groningen\\
The Netherlands\\
\2ad Theoretische Natuurkunde\\
Vrije Universiteit Brussel\\
Pleinlaan 2\\
B-1050 Brussels, Belgium

}
\def\abstracttext{We review an iterative 
construction of the supersymmetric non-abelian 
Born-Infeld action. We obtain the action through 
second order in the fieldstrength. 
Kappa-invariance fixes the ordenings which turn out to deviate from 
the symmetrized trace proposal.
}
\large
\makefront
\section{Introduction}

On a D$p$-brane there are eight propagating fermionic and eight propagating
bosonic worldvolume degrees of freedom. In the static gauge, the bosonic
ones appear as a $U(1)$ vector field in $p+1$ dimensions and $9-p$ scalar
fields. The former describe open strings attached to the brane while the
latter describe the transversal positions of the brane. The effective 
action for slowly varying fields is known through all orders in $\alpha '$,
it is the ten-dimensional $U(1)$ Born-Infeld theory dimensionally reduced to 
$p+1$ dimensions \cite{AT}, \cite{FT}. Its fully covariant 
supersymmetric generalization is known as well \cite{Aga}, \cite{Ced}, 
\cite{BT}. 
It contains\footnote{
We denote by $\mu,\nu=0,\ldots,9$ the spacetime indices and by
$i,j=0,\ldots,p$ the worldvolume coordinates $\sigma^i$.} the
embedding coordinates $X^\mu(\sigma)$ (of which, because of the worldvolume
reparametrization invariance, only the transverse
coordinates are physical degrees of freedom), a vector
field $V_i(\sigma)$, and an $N=2$ spacetime fermionic field
$\theta(\sigma)$.
Because of the presence of a local fermionic symmetry, called 
$\kappa$-symmetry, half of the fermionic fields 
can be gauged away. This leaves, in the static gauge, the fieldcontent 
of a $U(1)$ supersymmetric Yang-Mills theory in $p+1$ dimensions
which indeed describes $8+8$ degrees of freedom.

Once several, say $n$, 
D$p$-branes coincide, additional massless states appear enhancing the 
$U(1)^n$ gauge symmetry to a full $U(n)$ gauge symmetry \cite{EW1}. 
Both the fermions 
and the scalars transform now in the adjoint representation of $U(n)$.
Already at the purely bosonic level there are difficulties in defining a 
non-abelian generalization of the Born-Infeld action. 
They arise from the 
fact that the notion of acceleration terms becomes ambigous as can be 
seen from
\begin{eqnarray}
D_i D_j F_{kl}=\frac 1 2 \{D_i , D_j\}F_{kl}-\frac i 2 {[}F_{ij}, F_{kl}
{]}.
\end{eqnarray}
Based on the results of a direct calculation of the action through order 
$F^4$ 
\cite{GW}, \cite{BP}, and
assuming all terms proportional to anti-symmetrized products 
of fieldstrengths to 
be acceleration terms which are then ignored, a proposal was formulated 
for the non-abelian action \cite{AT2}.
The action assumes a form similar to the abelian case but, upon 
expanding it 
in powers of the fieldstrength, one first 
symmetrizes over all fieldstrengths and subsequently one performs 
the group theoretical trace. Other possibilities are discussed 
in \cite{AN}. However, 
by calculating the mass spectrum from the effective action 
in the presence of certain background fields and by comparing it
to the spectrum predicted by string theory, it can be shown that the 
symmetrized trace 
proposal is flawed from order $F^6$ on \cite{HT}, \cite{DST}.

As a direct calculation of the effective action at higher orders in 
$\alpha '$ seems out of reach, different approaches are called for. One 
possibility would be to use the mass spectrum as a guideline \cite{STT}. 
In the 
present paper we explore a suggestion in \cite{DST} and use 
$\kappa$-symmetry 
to fix the effective action. Full technical details can be found in 
\cite{BDS}. 

\section{The abelian case}

In order to avoid additional complications coming from the 
presence of transversal coordinates, we
focus throughout this paper on D9-branes. Lower dimensional branes can 
then be studied upon performing a suitable T-duality transformation \cite{TR}. 
The $\kappa$-symmetric lagrangian has the schematic form 
\begin{equation}\label{BIA}
{\cal L} = - e^{-\phi} \sqrt {-{\rm det}\,
\big (g+{\cal F}\big )}\ +\ Ce^{\cal F}\, ,
\end{equation}
with ${\cal F} = 2dV +B$.
The first term is the Born-Infeld lagrangian ${\cal L}_{\rm BI}$ while the 
second is the Wess-Zumino term ${\cal L}_{\rm WZ}$.
Both the NS-NS background fields $g,\,\phi,\,B$ and
all the R-R background fields $C$ are superfields, i.e., they are functions
of the superspace coordinates $(X^\mu,\theta)$. 
The $\kappa$-symmetry acts on the fermions as
\begin{equation}
\label{dt}
 \delta\bar\theta(\sigma) = \bar \eta (\sigma) \equiv
\bar\kappa(\sigma)\,(1+\Gamma)\,,
\end{equation}
where $\Gamma$ depends on worldvolume and background fields. 
The variation of the D-brane action is then
\begin{equation}
\label{varkappa}
  \delta{\cal L} = - \bar\eta\, (1-\Gamma) {\cal T}\,,
\end{equation}
where $\cal T$ is some expression in terms of the worldvolume and
background fields. Invariance is obtained provided $\Gamma$ satisfies
\begin{equation}
  \Gamma^2 = {\bf 1} \,.\label{g2}
\end{equation}
As can be shown by combining eqs. (\ref{dt}) and (\ref{g2}), 
and by using the fact that $\tr \Gamma=0$, we can use the 
$\kappa$-gauge invariance to eliminate half of the fermions. 
Working in a flat gravitational background and taking all other 
bulk backgroundfields to be
zero, we can use the $\kappa$-symmetry to put 
\begin{eqnarray}
\theta = \pmatrix{\theta_1\equiv \chi\cr
                                   \theta_2 = 0}\, .
\end{eqnarray}
Fixing the worldvolume reparametrizations 
by taking $X^\mu = \delta^{i\mu}\sigma^i$, one finds that the
complete Wess-Zumino term
vanishes and the Born-Infeld lagrangian
is given by \cite{Aga}
\begin{equation}
{\cal L}_{\rm BI} = - \sqrt {- {\rm det}\, \big (
\eta_{\mu\nu} + F_{\mu\nu} + \bar\chi\Gamma_\mu\partial_\nu\chi +
{\textstyle{1\over 4}}\bar\chi \Gamma^a\partial_\mu\chi \bar\chi\Gamma_a
\partial_\nu\chi\big )}\,\, .
\end{equation}
This lagrangian has 16 linear and 16 nonlinear supersymmetries which
are obtained from the original 32 supersymmetries of the N=2 superspace
after implementing the fact that they get deformed with
a field-dependent kappa-transformation upon fixing the kappa-gauge.

Taking a closer look at the origin of various terms in 
eq. (\ref{varkappa}),
one finds that the $\bar\eta\Gamma{\cal T}$ term  
arises from the variation of the Wess-Zumino lagrangian ${\cal L}_{\rm WZ}$,
while the 
$-\bar\eta{\cal T}$ originates from the variation of the Born-Infeld 
lagrangian ${\cal L}_{\rm BI}$. Expanding both $\Gamma$ and ${\cal T}$ in 
powers of the fieldstrength $F$,
\begin{eqnarray}
\Gamma=\sum_{i\geq 0}\Gamma_i\, ,\quad {\cal T}=\sum_{i\geq 0} 
{\cal T}_i,
\end{eqnarray}
we get
\begin{eqnarray}
\delta{\cal L}_{\rm WZ}&=&\bar\eta\Gamma_0{\cal T}_0 +\bar \eta 
\Gamma_0{\cal T}_1+\bar\eta\Gamma_1{\cal T}_0
 + \ldots,\label{v1} \\
\delta{\cal L}_{\rm BI}&=&-\bar\eta{\cal T}_0-\bar \eta 
{\cal T}_1+ \ldots,\label{v2}
\end{eqnarray}
where eq. (\ref{g2}) implies additional restrictions
\begin{eqnarray}
\left(\Gamma_0\right)^2={\bf 1}\, ,\quad \left\{
\Gamma_0,\Gamma_1\right\}=0,\ldots\label{v3}
\end{eqnarray}
This structure suggests an iterative procedure for obtaining the D-brane 
action. Because of its topological nature, the form of the Wess-Zumino 
term is severely constrained. It is itself given as an expansion in powers 
of the fieldstrength $F$. Varying the term independent of $F$ in it gives 
the first term in eq. (\ref{v1}). Combining this with the first relation 
in eq. (\ref{v3}), both $\Gamma_0$ and ${\cal T}_0$ get 
identified. Integrating ${\cal T}_0$ then yields the lowest order in 
$F$ of the Born-Infeld lagrangian ${\cal L}_{BI}$. Proceeding like this 
order by order in the fieldstrength $F$, one fixes ambiguities in 
${\cal L}_{WZ}$ and one constructs the Born-Infeld lagrangian ${\cal 
L}_{BI}$.

In the next section, we will adopt this strategy in order to obtain the 
non-abelian Born-Infeld lagrangian. 

\section{The non-abelian case}

When constructing the supersymmetric D9-brane action, one can aim for 
several goals. In order of increasing ambition they are given by:
\begin{enumerate}
\item The construction of the supersymmetric non-abelian Born-Infeld 
action with trivial IIB supergravity backgrounds in the static gauge. 
Besides the vector fields $V_i(\sigma)$, there is an $N=1$ spinor field 
$\chi$ in the adjoint of $U(n)$.
\item Repeat step 1, but now with a manifest worldvolume reparametrization 
invariance. In addition to the fields listed above, we have now the 
embedding coordinates $X^\mu (\sigma)$ as well.
\item Repeat step 1, but making the action invariant under 
$\kappa$-symmetry. Instead of the $N=1$ spinor $\chi$ we get now an $N=2$ 
spinor $\theta$ transforming in the adjoint of $U(n)$.
\item Combine the programmes listed under 2 and 3.
\item Repeat the previous programme in a background of 
non-trivial IIB supergravity bulkfields.
\end{enumerate}
Some initial steps towards achieving point 1 were made in \cite{CF} 
in four dimensions. In step 2 one has 
to decide whether the worldvolume 
embedding coordinates are singlets under $U(n)$ or 
whether they transform, in analogy with transversal coordinates, in the 
adjoint representation. The analysis in \cite{BDS} suggests that only 
the latter option is possibly consistent. However then another fundamental 
problem shows up. In order to be able to reach the static gauge, the 
structure of the worldvolume has to be adapted such as to obtain 
a sufficiently large reparametrization group. In order to proceed we opted 
for the programme in step 3 as steps 4 and 5 are presently out of reach.

We introduce fields $\theta^A(\sigma)$ \footnote{The 
$U(n)$ generators $T_A$, $A\in\{1,\cdots,n^2\}$, 
are Hermitian $n\times n$ matrices normalized
as $\tr T_AT_B = \delta_{AB}\,$.
The product of two $U(n)$ generators is given by
$T_AT_B = ( d_{ABC} + if_{ABC})T_C\,$, 
where $d$ and $f$ are symmetric and antisymmetric in $AB$, respectively.},
which are an $(N=2)$ doublet
of Majorana-Weyl spinors for each $A$,
satisfying $\Gamma_{11}\theta^A=\theta^A$. They transform as follows
under supersymmetry ($\epsilon$), $\kappa$-symmetry ($\kappa$) and
Yang-Mills transformations $(\Lambda^A)$:
\begin{equation}
\label{dtall}
   \delta\bar\theta^A(\sigma) = -\bar\epsilon^A
     + \bar\kappa^B(\sigma)({\bf1}\,\delta^{BA} +
        \Gamma^{BA}(\sigma))
     + f^A{}_{BC}\Lambda^B(\sigma)\bar\theta^C(\sigma) \,.
\end{equation}
Here $\epsilon^A$ are constant, $\Gamma^{AB}$ depends on the
worldvolume fields and it must satisfy
\begin{equation}
  \Gamma^{AB}\Gamma^{BC} = \delta^{AC}{\bf 1}\,. \label{gg2}
\end{equation}
{From} now on we use $\bar\eta^A \equiv  \bar\kappa^B(\sigma)({\bf 1}\,
 \delta^{BA} +
        \Gamma^{BA}(\sigma))\,$.
Because $\epsilon^A$ is constant we find from the commutator of
Yang-Mills and supersymmetry transformations that
$ f_{ABC}\epsilon^C=0$.
Therefore $\epsilon = \epsilon^A T_A$ must be proportional to the unit
matrix, i.e., there is only
one nonvanishing $\epsilon$ parameter.
Only after $\kappa$-gauge fixing will all $\theta$'s transform under
supersymmetry. The commutator of $\kappa$-symmetry and
supersymmetry teaches us that $\Gamma^{AB}$ is a supersymmetry invariant. 

We have now the tools at hand to start the programme outlined at the end 
of the previous section. We present the results and refer to \cite{BDS} for 
details of the calculation. Through second order in $F$,
the lagrangian is ${\cal L}={\cal L_{{\rm WZ}}}+{\cal L_{{\rm BI}}}$,
with the Wess-Zumino term given by
\begin{eqnarray}
    {\cal L}_{{\rm WZ}} &=& \epsilon^{i_1\ldots i_{10}} \,
     \bigg\{ \, {1\over 2\cdot 9!}\,\bar\theta^A\sigma_1
   \gamma_{i_1\ldots i_9}\Dpartial_{i_{10}}\theta^A
   \nonumber\\
   &&\quad - {1\over 4\cdot 7!}\,\bar\theta^A \, {\cal P}^{ABC}_{(1)}
      \gamma_{i_1\ldots i_{7}}\Dpartial_{i_{8}}\theta^B
       F^C_{i_9 i_{10}}
   \nonumber\\
\label{LWZall}
   &&\quad\quad + {1\over 16\cdot 5!}\,
   \bar\theta^A \, \big( - \sigma_1 {\cal S}^{ABCD} \big)
      \gamma_{i_1\ldots i_{5}}\Dpartial_{i_{6}}\theta^B
       (F^C F^D)_{i_{7}\ldots i_{10}} \bigg\}\,,
\end{eqnarray}
and the Born-Infeld lagrangian by,
\begin{eqnarray}
    {\cal L}_{{\rm BI}} &=& -\,\bigg\{
    1 + \half\,\bar\theta^A\gamma^i\Dpartial_i\theta^A
       - \half\, \bar\theta^A \sigma_1 {\cal P}^{ABC}_{(1)}
       \gamma_{[i}\Dpartial_{j]}\theta^B\, F^{ij\,C}
     \nonumber\\
     &&\quad  + \noverm{1}{4} F^{ij\,A} F_{ij}^A
       - \noverm{1}{2} \bar\theta^A
       {\cal S}_{ABCD} \gamma_{(i}\Dpartial_{j)}\theta^B
           \{F^{ik\,C}F_{k}{}^{j\,D} + \quart \eta^{ij} F_{kl}^CF^{kl\,D}\}
     \nonumber\\
\label{LBIall}
     &&\quad    + \quart\bar\theta^A  {\cal A}^{ABCD}
                      \gamma_{ijk}
      \{\Dpartial^k\theta^B F^{il\,C} F_l{}^{j\,D}
         - \Dpartial_l\theta^B F^{ij\,C}F^{kl\,D}\} \,\bigg\} \,,
\end{eqnarray}
where,
\begin{eqnarray}
\label{Pall}
    {\cal P}^{ABC}_{(1)} &=& (i\sigma_2)d^{ABC} \,,
    \\
\label{Sall}
    {\cal S}^{ABCD} &=& {\cal P}^{AE(C}_{(1)}{\cal P}^{BD)E}_{(1)}
                     =  -  d^{AE(C}d^{BD)E}\,,
    \\
\label{Aall}
    {\cal A}^{ABCD} &=&  {\cal P}^{AE[C}_{(1)}{\cal P}^{BD]E}_{(1)}
                      =  -d^{AE[C}d^{BD]E}\,.
\end{eqnarray}
The global supersymmetry 
and local $\kappa$-transformations are given by,
\begin{eqnarray}
\label{dthall}
   \delta\bar\theta^A &=& - \bar\epsilon^A+\bar\eta^A\,,
   \\
\label{dVall}
   \delta V_i^A &=& \half\,
      (\bar\epsilon^B+\bar\eta^B)\,\sigma_1{\cal P}^{BCA}_{(1)}
        \gamma_i\theta^C
      +\half\,(\bar\epsilon^B+\bar\eta^B)\,
            {\cal S}^{BCDA} \gamma_k\theta^C F^{ki\,D}
      \nonumber\\
    &&\quad
     +\quart\,(\bar\epsilon^B+\bar\eta^B)\,{\cal A}^{BCDA}\gamma_{ikl}\theta^C
                F^{kl\,D}\,,
\end{eqnarray}
where as mentioned before, $\epsilon^A$ satisfies
$f_{ABC}\,\epsilon^C = 0$.
Finally, $\Gamma^{AB}$ which was introduced in eq. (\ref{dtall}) is given by,
\begin{eqnarray}
  \Gamma^{AB} &=& \Gamma^{(0)}\,\bigg\{
     \sigma_1\delta^{AB}  + {\cal P}^{ABC}_{(1)} \half\gamma^{kl} {F}_{kl}^C
   \nonumber\\
  &&
   - \sigma_1 \,
    {\cal S}^{ABCD} \big( \noverm{1}{8} \gamma_{ijkl}
          F^{ij\,C} F^{kl\,D} -\quart F_{kl}^C F^{kl\,D}  \big) 
\label{Gamall}
     -\sigma_1\, {\cal A}^{ABCD}
           \half \gamma_{ij} F^{ik\,C} F_k{}^{j\,D} \bigg\}\,.
\end{eqnarray}

We proceed with fixing the $\kappa$-symmetry. Writing out the $N=2$ doublets
explicitly,
\begin{equation}
\Gamma = \left(
  \begin{array}{cc}
     0 & \gamma \\
     \tilde\gamma & 0
  \end{array}   \right)\,,
\end{equation}
we find that eq. (\ref{gg2}) implies $\gamma\tilde\gamma=
\tilde\gamma\gamma={\bf 1}$. 
Here $\gamma$, $\tilde\gamma$ are $32\times 32$ matrices, with
in addition indices $AB$, where $A,B$  run from $1$ to $n^2$.
Separating the fermions into $N=1$ fermions we get
for eq. (\ref{dthall}),
\begin{eqnarray}
  \delta\bar\theta_1^A &=& -\bar\epsilon_1^A + \bar\eta^A_1
       \,,\qquad
 \delta\bar\theta^A_2 = -\bar\epsilon_2^A + \bar\eta^A_2
            \,,  \label{gg3}
\end{eqnarray}
where using the relation between $\eta$ and $\kappa$ we get,
\begin{eqnarray}
  \bar\eta &=& \left(\bar\eta_1\quad \bar\eta_2 \right) =
       \left( \bar\kappa_1+\bar\kappa_2\tilde\gamma
                 \quad
                     \bar\kappa_2+\bar\kappa_1\gamma
                     \right)\,.
\end{eqnarray}
Using the $\kappa$-symmetry, we can put $\bar\theta_2=0$\,.
As a consequence $\kappa_2$ is fixed,
$  \bar\kappa_2 = \bar\epsilon_2 - \bar\kappa_1\gamma \,$.
Combining this with eqs. (\ref{gg3}) and (\ref{Gamall}), we obtain the 
supersymmetry transformations of the fermions,
\begin{eqnarray}
   \delta\bar\chi^A &=& -\bar\epsilon_1^A - \bar\epsilon_2^A
    + \bar\epsilon_2^B\,
    \{  d^{BAC}\,\half \gamma^{kl} F_{kl}^C
    \nonumber\\
     &&\quad    +\ {\cal S}^{BACD}\,
               ( \noverm{1}{8} \gamma_{ijkl}
          F^{ij\,C} F^{kl\,D} -\quart F_{kl}^C F^{kl\,D})
          \,\}   \,,
\end{eqnarray}
where we called $\chi^A\equiv \theta_1^A$. Implementing the gauge choice in 
eq. (\ref{dVall}), we get the transformation rules for the gauge fields as 
well,
\begin{eqnarray}
   \delta V_i^A &=&  -\half\, (\bar\epsilon_1^B - \bar\epsilon_2^B)
            d^{BCA}\gamma_i\chi^C
         -\quart \bar\epsilon_2^B
         d^{BED}d^{ECA} \gamma_{kl}\gamma_i\chi^C F^{kl\,D}
                 \nonumber\\
      &&\quad +\ \half (\bar\epsilon_1^B-\bar\epsilon^B_2)
            S^{BCDA} \gamma_k \chi^C F^{ki\,D}
        \,.
\end{eqnarray}
After gauge fixing, the Wess-Zumino term vanishes
since it was off-diagonal in the fermions $\theta_1$
and $\theta_2$. The Born-Infeld term is given by
\begin{eqnarray}
    {\cal L}_{{\rm BI}} &=& - \bigg\{
    1 + \half\,\bar\chi^A\gamma^i\Dpartial_i\chi^A
       + \half\, {d}_{ABC}  \bar\chi^A
       \gamma_{[i}\Dpartial_{j]}\chi^B\, F^{ij\,C}
       + \noverm{1}{4} F^{ij\,A} F_{ij}^A
\nonumber\\
\label{LBIgf}
     &&\quad
       +\ \noverm{1}{2}
       d^{AEC}d^{BDE}\, \bar\chi^A \gamma_{(i}\Dpartial_{j)}\chi^B
           \{F^{ik\,C}F_{k}{}^{j\,D} + \quart \eta^{ij} F_{kl}^CF^{kl\,D}\}
     \\
     &&\quad    -\ \quart  d^{AE[C}d^{BD]E}\, \bar\chi^A
                      \gamma_{ijk}
      \{\Dpartial^k\chi^B F^{il\,C} F_l{}^{j\,D}
         - \Dpartial_l\chi^B F^{ij\,C}F^{kl\,D}\} \,\bigg\} \,. \nonumber
\end{eqnarray}
It is clear that the
terms of the form $\bar\chi\partial\chi F^2$ are
not symmetric traces of $U(n)$ generators. The symmetric trace is given 
by,
\begin{equation}
  \tr T_{(A}T_BT_CT_{D)} =
    \noverm{1}{3}\,( d_{ABE}d_{CDE} + d_{CAE}d_{BDE} + d_{BCE}d_{ADE})\,,
\end{equation}
while the second line in (\ref{LBIgf}) contains only two of the three contributions
needed for the symmetric trace, the last line contains explicit
anti-symmetrizations
and can be rewritten in terms of structure constants,
\begin{equation}
\label{id3}
   d_{AEC}d_{BDE}-d_{AED}d_{BCE} = f_{ABE}f_{CDE}\,.
\end{equation}

\section{Conclusions}

In this paper we have obtained the non-abelian generalization of the
Born-Infeld action up to terms
quartic in the Yang-Mills field strength, and including all fermion
bilinear terms up to terms cubic in the field strength.
The terms of the form $\bar\chi\partial\chi F^2$ deviate from
the symmetric trace conjecture. 
The precise structure of the non-abelian Born-Infeld action remains an 
enigma. One clue is provided by the fact that in the
abelian case $\Gamma$ factorizes into a part that is polynomial in $F$, and
the inverse of the Born-Infeld action, which expands to an infinite
series in $F$. While such a factorization will be more complicated 
in the non-abelian
case \cite{BDS}, we need to pursue this programme to higher order in the 
fieldstrength \cite{BRS2} in order to see some pattern appearing. 
In addition, having the supersymmetric Born-Infeld at higher order, 
would allow us to study non-abelian BPS states.

The simplest of non-abelian BPS configurations arises as follows \cite{BDL}. 
Taking two D$p$-branes in the $(2,4,\cdots, 2p)$ directions,
we keep one of them fixed and rotate the other one subsequently
over an angle $\phi_1$ in the (2\,3) plane, over an angle $\phi_2$ in the 
(4\,5) plane, ..., over an angle $\phi_p$ in the $(2p\,2p+1)$ plane. The 
following table summarizes for various values of $p$ the BPS conditions 
on the angles (which are different from zero)
and the number of remaining supercharges.

\begin{center}
\begin{tabular}{|c|l|c|}\hline\hline
$p$ &BPS condition & susy's\\ \hline\hline
2 &$\phi_1=\phi_2$ & 8\\ \hline
3 &$\phi_1=\phi_2+\phi_3$ & 4 \\ \hline
4 &$\phi_1=\phi_2+\phi_3+\phi_4$ & 2\\ \cline{2-3}
  &$\phi_1=\phi_2$, $\phi_3=\phi_4$&4 \\ \cline{2-3}
  &$\phi_1=\phi_2=\phi_3=\phi_4$&6 
\\ \hline\hline
\end{tabular}
\end{center}

T-dualizing in the $3, 5, ..., 2p+1$ directions yields
two coinciding D$2p$ branes with magnetic fields, $F_{2i\, 2i+1}$, 
$i\in\{1,\cdots,p\}$, 
turned on. In the simplest case we have $ F_{2i\, 2i+1}\equiv f_i 
\sigma_3$, with $f_i$ constant. The relation between magnetic fields and 
angles is $ \tan(\phi_i/2)=2\pi\alpha 'f_i$. 
Translating the BPS conditions on the angles in 
conditions on the fieldstrengths, we get,

\begin{center}
\begin{tabular}{|c|l|l|}\hline\hline
$p$ &BPS condition &fieldstrengths\\ \hline\hline
2 &$\phi_1=\phi_2$ & $f_1=f_2$\\ \hline
3 &$\phi_1=\phi_2+\phi_3$ & $f_1=f_2+f_3+(2\pi\alpha ')^2f_1f_2f_3$ \\ \hline
4 &$\phi_1=\phi_2+\phi_3+\phi_4$ & $f_1=f_2+f_3+f_4+(2\pi\alpha')^2
(f_1f_2f_3+f_1f_3f_4$\\
&&$\qquad +f_1f_2f_4-f_2f_3f_4)$
\\ \cline{2-3}
  &$\phi_1=\phi_2$, $\phi_3=\phi_4$&$f_1=f_2$, $f_3=f_4$ \\ \cline{2-3}
  &$\phi_1=\phi_2=\phi_3=\phi_4$&$f_1=f_2=f_3=f_4$ 
\\ \hline\hline
\end{tabular}
\end{center}

In several cases we get $\alpha '{}^{2m}$ corrections. 
Therefore we will have to go at least to order $F^3$ in 
the supersymmetry transformation rules in order to
be able to compare our results to these predictions. 
In particular we will then also be able to analyze non-diagonal BPS 
configurations. As shown in \cite{BKOP}, 
the knowledge of $\Gamma$ is sufficient to elegantly perform this analysis.

Another intriguing point is the apparant incompatibility
between $\kappa$-symmetry and worldvolume reparametrisation invariance.
The analysis in \cite{BDS} suggests that also the worldvolume  
embedding coordinates transform in the adjoint of $U(n)$. 
 Needless to say 
a better understanding of this would have profound implications in 
the understanding of D-brane geometry and might facilitate the coupling to 
curved backgrounds. 

Finally, it would be interesting to investigate whether the superembedding
techniques developed in \cite{Bandos:1995zw} or the 
analysis of \cite{Bellucci:2000bd} can be applied to the problem at hand. 

\vskip0.5cm
\noindent
{\large \bf Acknowledgements}

\smallskip
\noindent
We like to thank
I.~Bandos,
M.~Cederwall,
S.~Ferrara,
S.F.~Hassan,
E.~Ivanov,
R.~Kallosh,
U.~Lindstr\"om,
C.~Nappi,
A.~Peet,
V.~Periwal,
D.~Sorokin,
J.~Troost and
A.~Tseytlin
for useful discussions. 
We are grateful to the Spinoza Institute, Utrecht, halfway
between Brussels and Groningen, for the hospitality extended to us.
This work is supported by the European Commission
RTN programme HPRN-CT-2000-00131, in which E.B. and M.d.R. are associated
to the university of Utrecht and A.S. is associated to the university of
Leuven.


\end{document}